\documentclass[12 pt, english]{article}

\usepackage{amsmath}
\usepackage{braket}
\usepackage{amssymb}
\usepackage{amsfonts}
\usepackage{latexsym}
\usepackage{graphicx}
\usepackage{latexsym}
\usepackage{color}
\usepackage{indentfirst}
\usepackage{hyperref}
\usepackage{verbatim}
\usepackage{multicol}
\usepackage{booktabs}
\usepackage{makecell}

\def\beq{\begin{eqnarray}}
\def\eeq{\end{eqnarray}}

\begin{document}

\begin{center}

{\Large\textbf{Land use/land cover dynamics on vulnerable regions in Uruguay approached by a method combining Maximum Entropy and Population Dynamics}}\\
\vspace{1cm}
\textbf{Johny Arteaga}$^{a,d}$ \footnote{e-mail address: \ johnyart@colostate.edu},  \textbf{Jhonny Agudelo}$^{b,d}$ \footnote{e-mail address: \ jhonny.agudelo@utec.edu.uy},
\textbf{Alejandro Brazeiro}$^{c}$\footnote{e-mail address: \ brazeiro@fcien.edu.uy}
\textbf{Hugo Fort}$^{d}$\footnote{e-mail address: \ hugo@fisica.edu.uy}\\
\vspace{0.5cm}
\small{\sl
{(a) USDA UV-B Monitoring and Research Program, Natural Resource Ecology Laboratory, Colorado State
University, Fort Collins, Colorado 80523 USA }\\
(b) Licenciatura en Tecnologías de la Información - LTI\\Grupo de Redes Complejas y Computación Ambiental - GRECO\\
Universidad Tecnológica del Uruguay - ITR Centro Sur\\
(c) {Facultad de Ciencias, Universidad de la República, Uruguay}\\
(d) {Grupo de Sistemas Complejos y Física Estadística, Instituto de F\'{i}sica\\Universidad de la República, 
Iguá 4225, 11400,  Montevideo -MVD- Uruguay}}

\end{center}



\vskip 6mm
\begin{quotation}
\noindent
{\large {\it Abstract}}.
\quad
We present an exploratory population dynamics approach, described by Lotka-Volterra (LV) generalized equations, to explain/predict the dynamics and competition between land use/land cover (LULC) classes over vulnerable regions in Uruguay. We use the Mapbiomas-Pampa dataset composed by 20 annual LULC maps from 2000-2019. From these LULC maps we extract the main LULC classes, their spatial distribution and the time series of areas covered for each class. The interaction coefficients between species are inferred through the pairwise maximum entropy (PME) method from the spatial covariance matrices for different training periods. The main finding is that this LVPME method globally outperforms the more traditional Markov chains approach at predicting the trajectories of areas of LULC classes.
\end{quotation}

\begin{description}
\item[keywords:] Land use/land cover; Pairwise Maximum Entropy modeling; Lotka-Volterra modeling; Markov chains; Mapbiomas-Pampa 

\newpage

\item[List of acronyms]:\\

LV: Lotka-Volterra.

LVPME: Method resulting from inferring the interaction matrix of the Lotka-Volterra equations through Pairwise Maximum-Entropy. 

LULC: land use/land cover.

MaxEnt: maximum-entropy. 

MAPE: mean absolute percentage error.

PME: pairwise maximum-entropy.

\end{description}
\newpage


\section{Introduction}\label{Sec1}
Land use/land cover (LULC) changes are catalogued as one of the main global change drivers. They reflect biodiversity loss caused by natural habitats transformations due to human activities \cite{foley2005global}\cite{diaz2019pervasive}\cite{sala2000global}\cite{pimm2000extinction}. In fact, it is estimated that changes in this kind of systems contribute to the current climate change \cite{gibbard2005climate}. \\
Thus, it is necessary to generate high resolution LULC maps to develop and use cost-effective approaches to identify and manage environmental risks. In this sense, the uses of high resolution imagery from  remote sensing is currently a fundamental tool for the study of spatial and temporal LULC changes \cite{ROGAN2004301}. In Latin America, the uses of remote sensing imagery has made it possible to detect LULC changes associated with natural forest being replaced by grazing grasslands and croplands \cite{graesser2015cropland}. \\
In the last two decades, the rising prices of commodities in the global market have driven the expansion of croplands over the Rio de la Plata basin,  (\textit{e.g.}, soy) and forest plantation development (\textit{e.g.}, eucalyptus), which have impacted the Rio de la Plata grasslands biome \cite{jobbagy2006forestacion}\cite{paruelo2006cambios}\cite{baldi2008land}\cite{modernel2016land}\cite{overbeck2007brazil}. Indeed, natural grasslands represent one of the most at risk biomes on the planet, due to their high rate of conversion to other uses and low protection level \cite{henwood2010toward}\cite{hoekstra2005confronting}.\\
Therefore, it is imperative to build robust models to forecast the complex LULC dynamics  to support climate predictions and planning vulnerable biomes protection policies \cite{baldi2008land}\cite{brazeiro2020agricultural}. A great diversity of LULC models have been developed, including agent-based models, artificial neural networks, cellular automata, economics-based models, Markov chains and statistical modeling via logistic regression (see for instance \cite{Lantman2011} and references therein). These models can also be categorized, depending on the amount of detail included, as either whole
landscape models, distributional landscape models, or spatial landscape models \cite{Baker89}\cite{lambin1997}).  Markov chain models, since their first application in LULC dynamics \cite{burnham_1973}, have been one of the most widely used to describe land-cover maps generated by remote-sensing image classification  \cite{kumar2013}.  In particular, this approach has been used over the Rio de la Plata biome \cite{baldi2008land}\cite{apellaniz2021temperate}. Another recent approach applied to protected areas in Uruguay is economics-based modeling \cite{brazeiro2020agricultural}. In addition, hybrid methods have been recently implemented coupling Markov chains with cellular automata or Markov chains with Multi-layer Neural Networks  \cite{ozturk2015urban}\cite{yang2012spatiotemporal}\cite{girma2022land}. \\
This work introduces a novel approach inspired in community ecology to describe the interactions and dynamics between classes from LULC maps  through Lotka-Volterra (LV) generalized equations \cite{pastor2008mathematical}\cite{fort2020ecological}. Our model considers each LULC class as a "species" competing for on single resource, the land area.  The strength of the  interactions between pairs of LULC classes are determined indirectly using {\it pairwise maximum entropy} (PME) modeling
\cite{10.1371/journal.pcbi.1004182} \cite{fort2020ecological}. The resulting method is denoted Lotka-Volterra Pairwise Maximum Entropy (LVPME) method.
The data we use are the annual remote-sensing classification maps of LULC in the period 2000-2019 for vulnerable regions in Uruguay, whose natural grasslands are at risk as a consequence of intensive agriculture and forest plantation activities.

\section{Materials and Methods }\label{Sec2}
\subsection{Study Area}
\begin{figure}[t!]
    \centering
    \includegraphics[scale=0.2]{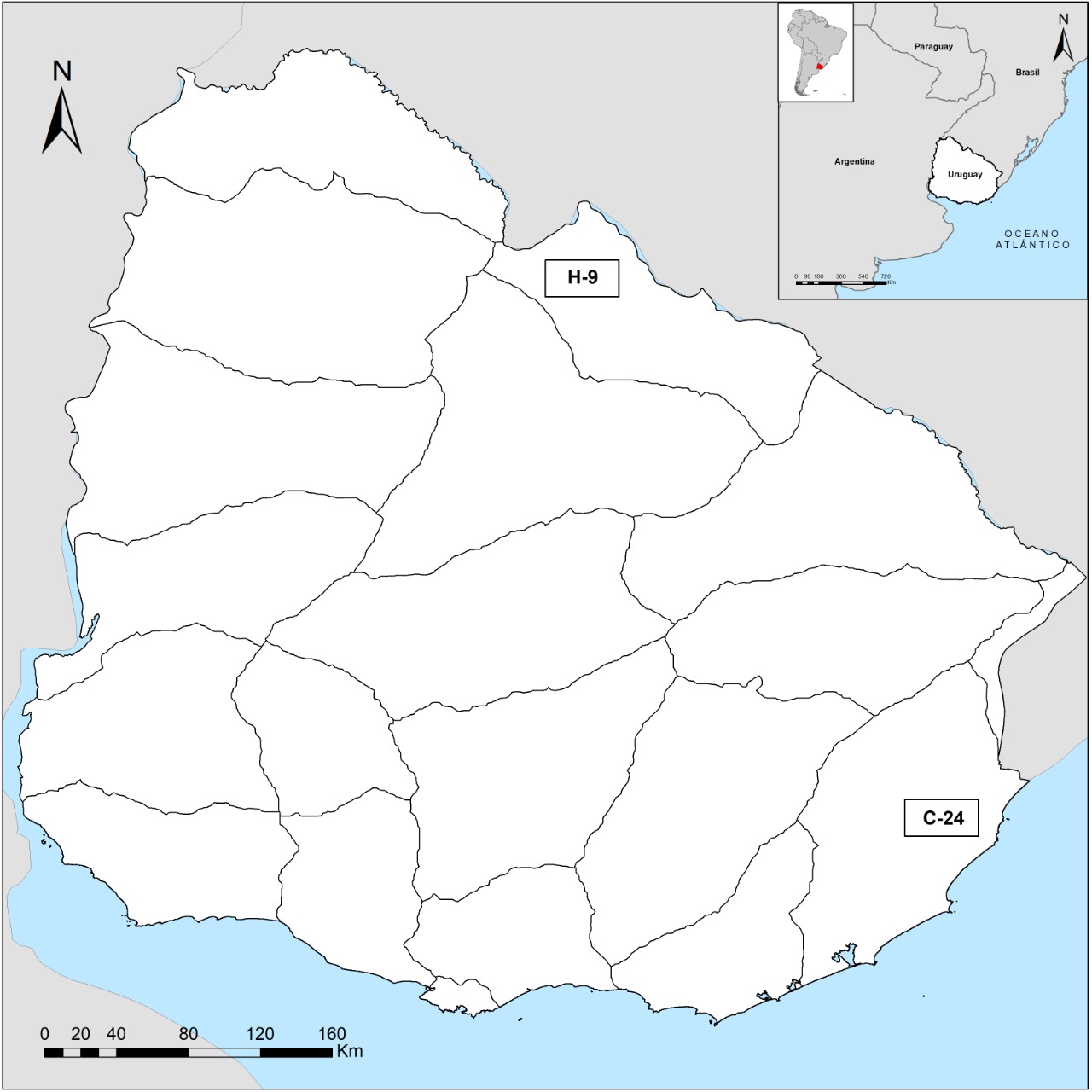}
    \caption{Regions of interest in Uruguay. H-9, to the north, it has been influenced by exotic forest plantation. C-24, to the east, it is a region influenced by agricultural activity.}
    \label{areas}
\end{figure}
Uruguay is a country located 
in southeast South America and its entire territory belongs to the Rio de la Plata grassland biome. The more relevant drivers influencing land use transformation are the croplands and exotic forest plantations \cite{brazeiro2020agricultural}. Figure \ref{areas} shows two regions of interest, denoted by code numbers C-24 and H-9 \cite{IGM}, each one  with an area of $6.6 \times 10^4 $ ha. The region H-9, located in the north of Uruguay, has been altered by exotic forest plantation since 1990, while the region C-24, located at the east of Uruguay, is distinguished by its intense agricultural activity. In order to analyze the LULC classes distribution and to estimate averages of abundances, covariances or interaction coefficients, both regions were divided into $1058$ quadrats measuring  $790 \times 790 $m approximately. 
\subsection{LULC dataset}

The recently created Mapbiomas-Pampa project, in its first collection, launched the annuals LULC maps from year 2000 to 2019 covering all the Pampa of Rio de La Plata \cite{mapbiomas}. In this work we use LULC maps starting at year 2000 and ending at year 2013, 2014,...,2017 as training sets and the remaining maps as validation sets (see Section \ref{predval}). In other words, the training period, $T_T$, varies from 14 to 18 years and the validation period, $T_V \equiv 20 -T_T $, varies from $2 - 6$ years.

\subsection{LV equations for the LULC system}\label{Sec4}
The growth of an isolated species is usually modeled in terms of the logistic equation, a model that relates the change in abundance or population density to the current abundance at time $t$, $A(t)$, its intrinsic growth rate, $r$, and its carrying capacity, $K$. The idea is that it seems reasonable that as the population density increases, the net growth slows down due to competition for resources \cite{pastor2008mathematical}. Thus, independently of the initial value of the abundance, $A(0)$,  $A(t)$ eventually converges asymptotically towards $K$. 
The LV competition equations are a generalization of the logistic equation, when the growth of a species depends, besides on the intraspecific interaction, on interspecific interactions. Using a discrete time formulation, these LV competition equations for $S$ interacting species can be written as the following set of finite difference equations for the abundance $A_i(t)$ of species $i$ at year $t$:
\begin{equation}
     A_i (t+1)-A_i (t)=r_i A_i(t)\left(1+\sum_{j}\frac{\alpha_{ij}A_j(t)}{K_i}\right)
     ,\qquad\qquad i=1,2,...,S,
         \label{set}
\end{equation}
where $r_i$ and $K_i$ are the same parameters of the logistic equation, i.e. the intrinsic growth rate and carrying capacity of species $i$, respectively; and $\alpha_{ij}$ are the coefficients of the interaction matrix \cite{pastor2008mathematical}\cite{fort2020ecological}. The magnitude of these coefficients is proportional to the influence of species $j$ over the growth rate of species $i$. In general,  the sign of these coefficients determines the nature of the interaction as
\begin{equation}
    \alpha_{ij}=\begin{cases}
  >0  & \textcolor{black}{facilitation} \\
  <0 &  \textcolor{black}{competition}
\end{cases}
\end{equation} 
Therefore, the larger the absolute value of the coefficient $\alpha_{ij}$, the larger is the facilitation or competition, and the influence of the species $j$ over species $i$ \cite{pastor2008mathematical}\cite{fort2020ecological}. 

In our case, the role of species is played by each LULC class, and the abundances of species correspond to the areas covered by each class, so that the total area is the finite resource they are competing for (see Table \ref{tabcon} for more details). \\
\subsection{Parameter estimation}\label{Sec31}

\begin{table}[t!]
\centering
\caption{\small Parameters description of Lotka-Volterra competition model for multiple interacting LULC classes.
}\vspace{0.3cm}
\begin{tabular}{ccc}
\toprule
 \thead{\textbf{LV parameter}}   & \textbf{\thead{Original meaning for multiple \\ interacting populations } }& \textbf{\thead{Adjusted meanings for multiple \\interacting LULC classes}}  \\
 \cline{1-3}
\thead{$A_i$} & \thead{Abundance of $i$-th specie } & \thead{Area occupied by the $i$-th LULC class}\\
\thead{$r_i$} & \thead{Intrinsic growth rate of the $i$-th specie, \\ given by the balance between \\natality and mortality} & \thead{Intrinsic growth rate of the $i$-th LULC class, \\ given by the balance between the rates \\of expansion and  retraction in the region.} \\
\thead{$K_i$} & \thead{Carrying capacity: asymptotic value of $A_i$ \\when specie $i$ is growing in isolation from \\ the other species, being only limited \\by resources availability.} & \thead{Carrying capacity: asymptotic value of $A_i$\\ when LULC is growing without restrictions \\imposed by other classes, being limited by \\the amount of suitable area in the region, \\and by economic or legal  restrictions \\in the case of anthropic LULCs \\ (e.g., agriculture, tree plantation, urbanization). } \\
\thead{$\alpha_{ij}$} & \thead{Interaction coefficient: strength of the \\interactions of specie $j$ over specie $i$, \\given by the competitive (or facilitation) \\ abilities of each species} &  \thead{Interaction coefficient: strength of the \\ interactions of LULC class $j$ over class $i$, \\ given by the competitive abilities of each \\ LULC class,  that could be regulated by \\ ecological  (e.g., species colonization rates) and \\ anthropogenic factors  }\\
\bottomrule
\end{tabular}
\label{tabcon}
\end{table}
We have to estimate from data three sets of parameters: $\{r_i\}$ $\{K_i\}$ and $\{\alpha_{ij}\}$. We will show firs how to infer $\{\alpha_{ij}\}$ through PME modeling, and, once we have this set of parameters, how $\{r_i\}$ and $\{K_i\}$ can be estimated by linear regression. 
\subsubsection{Maximum entropy method to quantify the strength of interactions}\label{Sec41}
PME modeling is based on the principle of Maximum Entropy (MaxEnt) \cite{jaynes1957information}\cite{jaynes1957information2}, a general method to make the least biased inferences compatible with available (incomplete) data. Specifically, to obtain the probability distribution which best represents our current state of knowledge about a system, the approach of MaxEnt is to  maximize the Shannon’s information entropy\cite{shannon1948mathematical} subject to constraints on expected values, i.e., averages, representing empirical observations. 
Thus, imagine a set of interacting entities, e.g., species, and that for each species we have a sample of several measurements of its abundance
(that can correspond to different times or sites or experimental replicas, etc.) from which we can compute sample averages. 
The simplest of such averages which can be included as constraints are the sample means. If one is interested in the pairwise 
interactions between these variables, the matrix of sample covariances $C_{ij}$ is the next simplest set of constraints that simultaneously 
takes into account their pairwise interactions 
and can be expressed as an average. The PME models are those resulting from taking the sample means and covariances as constraints\cite{stein2015inferring}. It is straightforward to show that, maximizing the information entropy with these constraints, the resulting probability distribution is a Gaussian involving in its exponent a matrix $J_{ij}$ connecting pairs of species $i$ and $j$ given by \cite{HFort}:  
\begin{equation}\label{Jij}
     \textcolor{black}{J_{ij}=-C_{ij}^{-1}}.
\end{equation}
The natural interpretation for the elements of matrix $J_{ij}$ is that they quantify the {\it effective} strength, i.e., of all the direct and indirect operating  effects, of species $j$ on the growth rate of species $i$ \cite{stein2015inferring}\cite{HFort}.
Indeed, interpreting this matrix as an interaction matrix has recently demonstrated to be useful in community ecology, for example to predict the dynamics of tree species in tropical forests \cite{fort2021method} and to generate early warnings of impending population crashes of tree endangered species \cite{fort2021new}. 
Therefore, for regions C-24 and H-9, we compute for a time $t$ the covariance matrix $C_{ij}(t)$ as 
\begin{equation}\label{Cij}
    C_{ij}(t)= \overline{{ (A_i - \bar{A}_i) (A_j - \bar{A}_j)}},
\end{equation}
where the bar denotes the average value over the classes. Since $C_{ij}$ depends on the year $t$ used to compute it, so it does $J_{ij}$; i.e. $J_{ij}=J_{ij}(t)$. 
In addition, notice that  $J_{ij}$ is a symmetric matrix because $C_{ij}$ is symmetric. 
 In fact, it was found that a better descriptor of the net interaction a species $j$ has on the growth of species $i$ is a normalized version of $J_{ij}$, obtained by dividing each row $i$ of $J_{ij}$ by the absolute value of the diagonal element $J_{ii}$  \cite{fort2020ecological,emary2021markets}. Therefore, here we take:
\begin{equation}\label{alphaij}
    \alpha_{ij}(t)=\frac{J_{ij}(t)}{|J_{ii}(t)|},
\end{equation}
which is asymmetric and thus more realistic for describing arbitrary interactions in nature (as the one existing between LULC classes).  
\subsubsection{Estimation of growth rates and carrying capacities}\label{Sec5}
Eq. (\ref{set}) can be rewritten as:
\begin{equation}
    \textcolor{black}{\hat{y}_i}=\frac{r_i}{K_i}\textcolor{black}{\hat{x}_i}+r_i
\end{equation}
where we have defined the quantities 
\begin{equation}
    \hat{y}_i\equiv\frac{A_i (t+1)-A_i (t)}{A_i (t)} ,\qquad\hat{x}_i\equiv\sum_{j}\alpha_{ij}A_j
    \label{rk}
\end{equation}
such that we find a simple linear relation between $\hat{y}_i$ and $\hat{x}_i$, therefore allowing us to estimate $r_i$ and $K_i$ using a linear regression \cite{fort2021new}. In this work, this linear regression will be performed considering different lengths of the training period, varying from $T_T$=14 to 18 years. For example, for $T_T$=14 years, the initial and final times are $t_i=2000$ and $t_f=2013$, respectively.

\subsection{Model validation}

Once the $r_i$ and $K_i$ were estimated, we use the LV competition equation to obtain  the forecasted trajectory of the area of each class, $F_i(t)$,  starting from $F_i(T_T) = A_i(T_T)$ and varying $t$ from $t=T_T+1$ to $t=T_V$, as:
\begin{equation}
F_i (t+1)=F_i (t)\left[1+r_i\left(1+\sum_{j}\frac{\alpha_{ij}F_j}{K_i}\right)\Delta t\right], \;\;\; t=T_T,\dots T_V-1. \label{lv-sol}
\end{equation}
To quantify the accuracy of predictions we consider three metrics. The first one is the mean absolute percentage error (MAPE) defined for each class as  
\begin{equation}
    \mbox{MAPE}_i= \frac{100}{T_V}\sum_{t=1}^{T_V} \left|
    \frac{A_i(t)-F_i(t)}{A_i(t)}\right|,
    \label{mape1}
\end{equation}
where $A_i(t)$ and $F_i(t)$ are the actual and forecasted values for time $t$, respectively.
It serves to assess the goodness of the trajectories predicted by the model for each species.
A second metric is provided by the percentage of area which is wrongly predicted for each year $t$ of the validation set, given by:
\begin{equation}
    \epsilon(t)=100\times\sum_{i=1}^{3}\frac{|A_i(t)-F_i(t)|}{\sum_{i=1}^{3}A_i(t)},  \;\;\; t=T_T+1,\dots T_V.
    \label{wrongarea}
\end{equation}
Finally, by taking the average over classes of MAPE$_i$, we obtain a third metric of global accuracy of the method given by
\begin{equation}
\overline{\mbox{MAPE}} = \frac{1}{S} \sum_{i=1}^S \mbox{MAPE}_i \label{mape2}
\end{equation}

\subsection{The benchmark: Markov chains}\label{Markov}
 Markov chains, because of its simplicity to describe LULC changes have been widely used to predict the spatio-temporal LULC dynamics  \cite{Lantman2011}\cite{Baker89}\cite{lambin1997}\cite{kumar2013}.
Hence, in this study Markov chains will be used as a benchmark to compare the performance of our method. 
 A simple Markov chain is described by the following equation
\begin{equation}
 \chi(t_f + 1) = P^T(t_i, t_f) \, \chi(t_f), \label{markov1}
\end{equation}
where $\chi(t_f)$ is a probabilistic state vector composed by the probability of occurrence of  each class $i$  at time $t_f$ given by
\begin{equation}
    \chi(t_f) = \{\chi_i(t_f)\} = \Bigl\{\frac{A_i(t_f)}{\sum_j A_j(t_f) } \Bigr\}. \label{markov2}
\end{equation}
$ P^T(t_i, t_f) $ is the transposed of the transition probability matrix computed from the initial time $t_i$ until the final time $t_f$ \cite{wang2021land}. The transition probability matrix $P(t-1, t)$ corresponds to the probability that the current class $i$ at year $t-1$ is replaced by the $j$ class in the consecutive year $t$. The transitions are computed superimposing a quadrat grid for two LULC maps at time $t-1$ and time $t$ and selecting the dominant LULC class in each of the $1058$ quadrats. Thus the matrix elements of $P(t-1, t)$  are given by
\begin{equation}
    P_{ij}(t-1, t) = \frac{N_{ij}(t)}{N_{i}(t-1)} \label{markov3}
\end{equation}
where $N_{ij}(t)$ is the number of quadrats dominated by class $i$ at year $t$ and later by class $j$ at year $t$, and $N_{i}(t-1)$ is the number of quadrats dominated by class $i$ at year $t-1$. 
Assuming the Markovian property, the multi-step Chapman-Kolmogorov equation is given by 
\begin{equation}
    P(t_i, t_f ) = P(t_i, t_{i} + 1) \cdot\, ...\,  \cdot P(t_{f}-1, t_f), \label{markov4}
\end{equation}
for the transition probability between $t_i$ and $t_f$ \cite{WENG2002273}. In this work initial time $t_i$ will be the 2000 year and the final time $t_f$ will vary according to the different training periods $T_T$.  


\section{Results and Discussion}\label{Res}
\begin{figure}[t!]
     \hspace{-2cm}\includegraphics[scale=1.0]{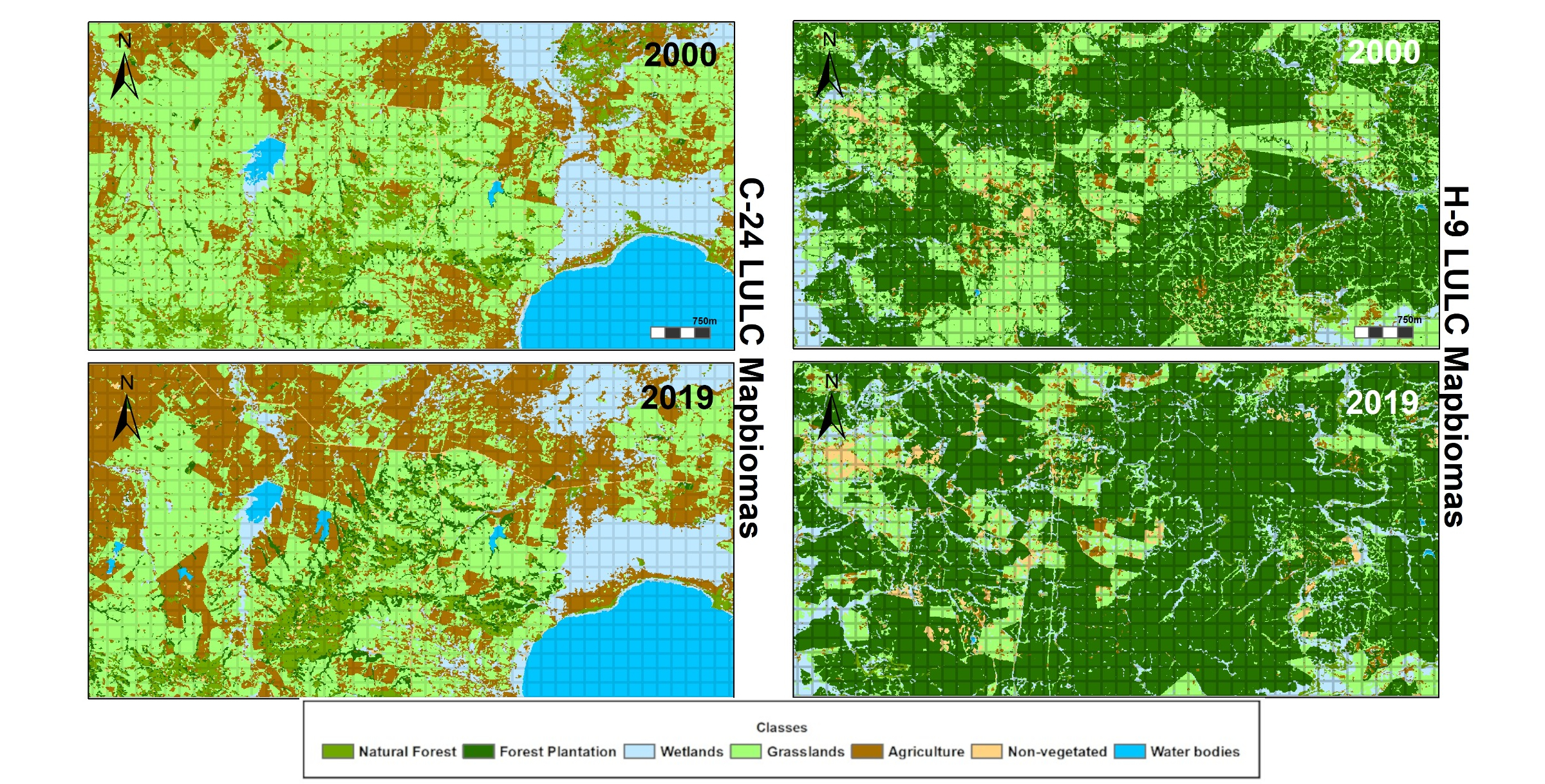}
    \caption{LULC maps of year 2000 and 2019 for the two areas of interest C-24 (left) and H-9 (right). Each map is divided by a gray grid formed by $1058$ quadrats measuring $790 \times 790 $m.}
    \label{areas-2}
\end{figure}

\begin{figure}[t!]
    \centering
    \includegraphics[scale=1.2]{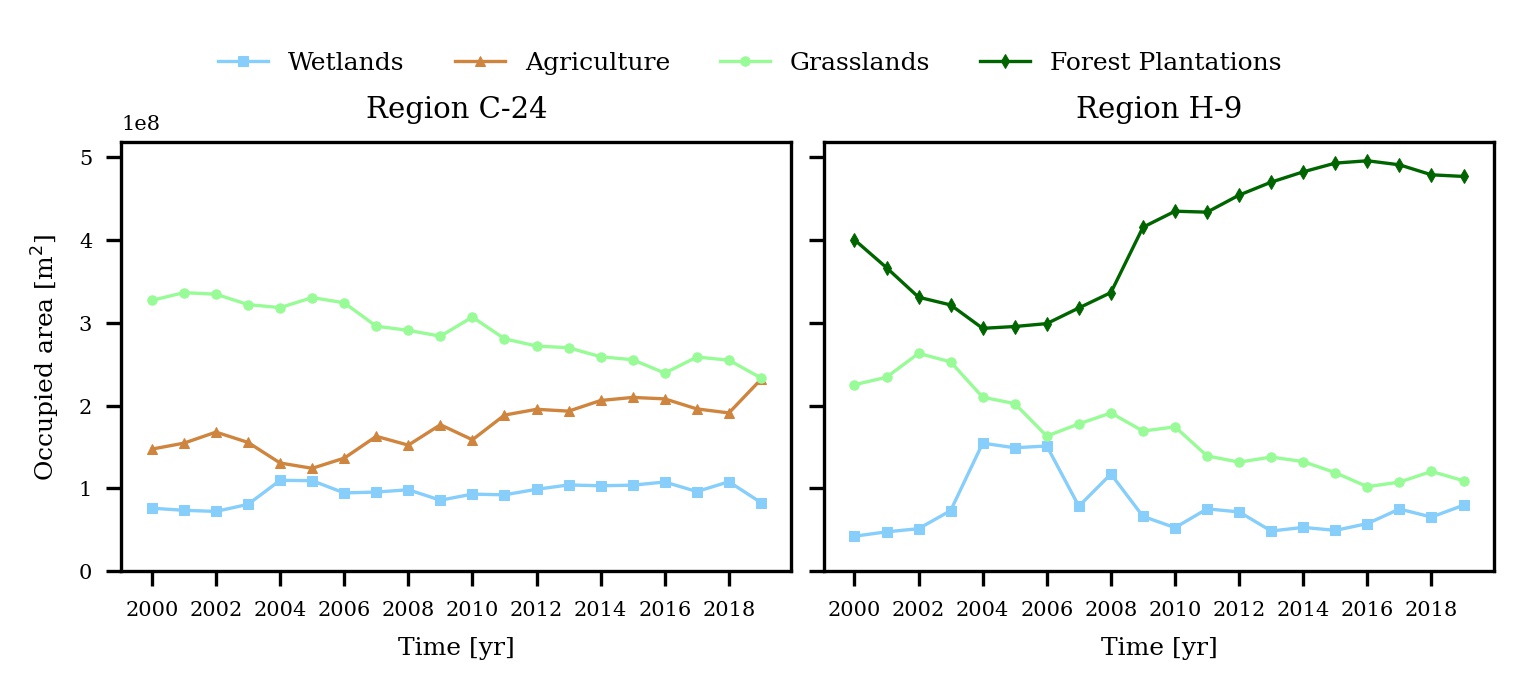}
    \caption{Observed trajectories of the area of each class,  $A_i(t)$, between 2000 and 2019 for the three selected classes from the Mapbiomas-Pampa database.}
    \label{timeseries}
\end{figure}

Fig. \ref{areas-2} are LULC maps generated by the Mapbiomas-Pampa  Project  for the years 2000 and 2019 in both regions of interest. We can observe for the region C-24 that the Grasslands, which dominated in 2000, have experienced a great decrease at the expenses of the Agriculture. In region H-9, Forest Plantation was the dominant class and has increased its dominance by displacing Grasslands. In both regions, we selected the classes with an occupied area above $10\%$ of the total. In the case of C-24 these classes correspond to Agriculture,  Grasslands and Wetlands, while for H-9, the classes are Forest Plantation, Grasslands and Wetlands. \\
Fig. \ref{timeseries} shows the time evolution of the occupied area for the three dominant classes (in m$^2$) for  C-24 and H-9 regions. Notice that the Grasslands have been losing occupied area in favour of Agriculture over C24 region and the Forest Plantations in H-9 region.  The C-24 region seems to be more stable while larger variations are observed in H-9, where the   Wetlands presents a high annual  variability  in the period 2003-2009, which may be related to a misclassification problem  in the reported  LULC maps due to the confusion between post-harvested or young forestry plantations with open areas like grasslands and wetlands from satellite imagery  and others issues recognizing these biomas \cite{wetlands}, \cite{wet2}. 
In addition, in region H-9 the decreasing observed in the period 2016 - 2019 in Forest Plantation  can be attributed to the continuous decaying in the  cellulose pulp price  during the  period 2010 - 2016 \cite{prices}, which reflects the  multi-agents dependency  of the LULC dynamics. 

\subsection{Growth rates and carrying capacities}

Having computed  the interaction matrices across time, we estimated the growth rates $r_i$ and carrying capacities $K_i$ using Eq. \eqref{rk}.  Fig. \ref{rk_c24} shows the $r_i$  and $K_i$ parameters  for C-24 and H-9 regions as a function of training set size. For C-24, we observe that the growth rates of Agriculture and Wetlands, $r_{A}$ and $r_{W}$, have a nearly constant behaviour in the 2015-2019 period, whereas the Grasslands growth rate $r_{G}$ present the highest value across time and greater variability in the 2017-2019 period. In H-9 region, despite decrease in growth rate $r_F$ of Forest Plantation, , it holds its dominant position over Wetlands and Grasslands.  However, it is worth remarking that what we observe for each class $i$ is the {\it effective} growth rate that results from the product of its intrinsic growth rate $r_i$ times the factor $\left(1+\sum_{j}\alpha_{ij}A_j(t)/{K_i}\right)$. \\
With respect to carrying capacities, in C-24, Grasslands holds a quasi stationary profile along  time in contrast to the Agriculture and Wetlands. In H-9 region, the carrying capacities for all classes are constant, where the Forest Plantations value is close to the total area for this region ($6.6 \times 10^8$ m$^2$), corresponding to the hypothetical value whereby each class completely dominates the area. Additionally in H-9, despite of the low variations in $r$ and $K$ values, the intrinsic growth associated with Wetlands class is slightly greater than Grasslands, in contrast to the opposite situation recorded for the time series for the abundances (see Fig. \ref{timeseries}).
\subsection{Validation}\label{predval}

Tables \ref{tab1} and \ref{tab2} show the  performance of predictions using the LV equations \eqref{lv-sol} and Markov chains \eqref{markov1}-\eqref{markov4} for different training periods of regions  C-24 and H-9, respectively. Note that, the matrix elements of transition probabilities are shown for different training years. For example, considering a training period between 2000-2013 years, the matrix $P(2000,2013)$ is given by
\begin{equation}
  \begin{array}{cccc}
       & Wetlands & Grasslands & Agriculture  \\
      Wetlands & 0.535 & 0.182 & 0.283  \\
      Grasslands & 0.069 & 0.635 & 0.296  \\
      Agriculture & 0.204 & 0.416 & 0.380  
  \end{array},  
\end{equation}
which satisfies the condition $\sum_{i}P_{ij}=1$. \\ 
For region C-24, despite that Table \ref{tab1} shows a similar MAPE between Markov chains and the LV method, we observe that LV method has a better performance only for the Wetlands class. In region H-9 the LV method is favoured over Markov chains for all classes. In addition, the average over classes MAPE in Table \ref{tab3} shows a better performance of LV over Markov chains for both regions and for all training years. This behavior contrast with the Markov domination in region C-24 considering each class separately (see again Table \ref{tab1}).

\subsection{Forecasting 2016 - 2025}

\begin{figure}[t!]
    \centering
    \includegraphics[scale=1.26]{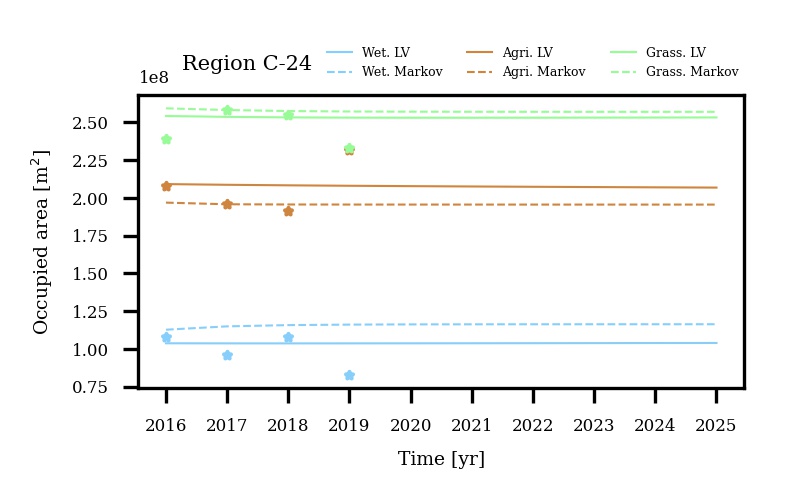} \hspace{-0.7cm}
    \includegraphics[scale=1.25]{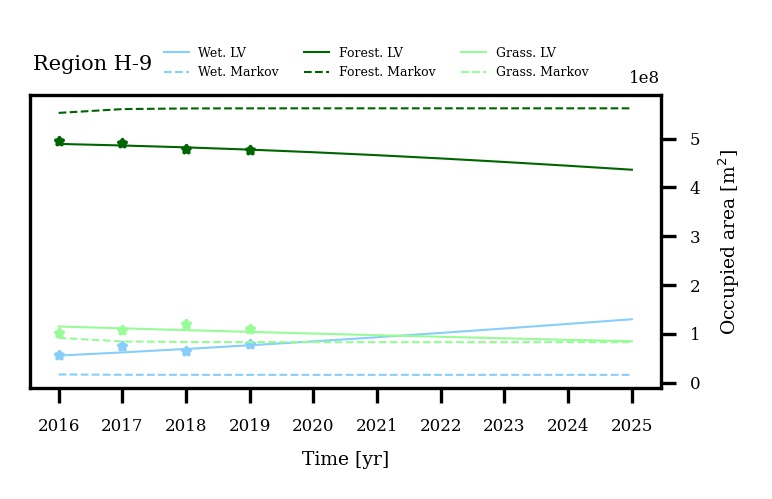}
    \caption{Ten years forecasting in the period 2016 - 2025  for the main LULC classes in regions C-24 and H-9 by LV method (solid lines) and simple Markov chains (dashed lines). The symbols represent the observed data points from Mapbiomas Pampa Project.  }
    \label{projec24}
\end{figure}

 From Table \ref{tab3} and Table \ref{tab4} is observe that the training period with less $\overline{\mbox{MAPE}}$ and $\epsilon$  corresponds to 2000-2015. Therefore  we selected this training period to forecast the next ten years (2016 - 2025).  Figure \ref{projec24} presents the goodness of predictions for LV and Markov methods in the period 2016-2025, while the  observed data  up to 2019 are presented by symbols.  The projections based on the LV model better fit the observations (2016-2019) of the different LULC classes in both regions. In C-24 region,
the performance of LV model is slightly better than Markov chain, that overestimate wetlands and underestimate agricultural areas. But both methods project stability in this region until 2025, with a dominance of grasslands over
agriculture and wetlands.
In the H-9 region, the fit of the LV model to the observations was much better than that of the Markov chain, giving more support to its projections. While the LV model projects slight increasing and decreasing trends for wetlands and forestation respectively, the Markov chain projects stability in all covers. These results suggest that LV model better
capture the trends observed in these LULC classes during 2016 and 2019, projecting them to the next six years. Markov chain model generates projections of future states based solely on its previous state (Eq. \ref{markov1}), so it is
reasonable this method has limitation to capture among years trends, and therefore to projects trends in the future.

We finish this section with a note of caution. A general caveat to all quantitative LULC modeling methods is that misclassification issues presented by maps built from satellite imagery introduce errors \cite{rs}.


 \section{Conclusions}\label{Conc}

  We have introduced a new method for describing LULC dynamics which is based in the deterministic  Lotka-Volterra model of competition between species represented by the LULC classes, whose parameters are described in Table \ref{tabcon}. The relationship between the  The interaction rules between  LULC classes  is estimated through PME modeling along a defined  training period. That is, PME modeling does not incorporate any explicit interaction mechanism between LULC classes. Instead, PME produces effective or net interaction coefficients that incorporate all the internal and external causes that lead to the spatial covariation of the areas covered by these classes. We call the resulting method from the combination of Lotka-Volterra + PME modeling, the LVPME method.

Of the three metrics considered to evaluate the model performance respect to the Markov chain benchmark:
 
 \begin{itemize}
 
 \item MAPE: The model clearly outperformed the benchmark in the more dynamic H-9 region while it unperformed in the more stable C-24 region.
 
 \item Wrongly predicted area: The model outperformed the benchmark in both regions. In C-24 the advantage in favour of the model is slight while in H-9 the model wins in most of the instances. 
 
 \item Mean MAPE:  The model outperformed the benchmark in both regions. 
     
 \end{itemize}
 
We thus conclude that the LVPME method is a useful tool for predicting the evolution of LULC classes by a set of finite  difference equations coupled by coefficients which take into account the spatial distribution of the occupied area by each class. This method shows to have a better performance than  traditional Markov chains, opening up the opportunity to complement standard  methods that perform  its predictions based on  spatial rules. 


\begin{table}[b!]
\centering
\caption{\small \textbf{C-24.} Multi-step transition matrix probabilities for different training periods and MAPE for validation until 2019 of the Markov chain benchmark and Lotka-Volterra (LV) modeling. Italic numbers are used for LV whenever it yields a larger MAPE than the Markovian benchmark.}\vspace{0.3cm}
\begin{tabular}{cccccccc}
\toprule
 Years   & LULC & \multicolumn{3}{c}{Transition probabilities $P_{ij}$ \eqref{markov4}} & &\multicolumn{2}{c}{MAPE} \\
\cline{3-5} \cline{7-8}
  t$_i$-T$_T$    &  & Wetlands & Grasslands & Agriculture & &Markov & LV  \\ 
\hline
2000-2013 & Wetlands & 0.535 & 0.182 & 0.283 & & 16.41 & 6.97
\\
2000-2014 &  & 0.513 & 0.188 & 0.299 && 13.75 & 8.53 \\
2000-2015 &  & 0.505 & 0.200 & 0.294 && 12.5 & 9.94\\
2000-2016 &  & 0.492 & 0.206 & 0.301 && 19.76 & 16.70 \\
2000-2017 & & 0.450 & 0.600 & 0.320 && 13.83 & \it{15.50}\\
2000-2013 & Grasslands & 0.069 & 0.635 & 0.296 && 5.34 & \it{9.15} \\ 
2000-2014 &  & 0.070 & 0.604 & 0.326 && 3.68 & \it{5.16} \\
2000-2015 &  & 0.073 & 0.596 & 0.331 && 3.2 & \it{4.36}  \\
2000-2016 & & 0.079& 0.570 & 0.351 && 4.85 & \it{5.08}\\
2000-2017 && 0.080 & 0.600 & 0.320 && 9.08 & 7.94 \\
2000-2013 & Agriculture & 0.204 & 0.416 & 0.380 && 13.45 & \it{15.83} \\
2000-2014 &  &  0.199 & 0.405 & 0.396 && 7.08 & \it{8.43}\\
2000-2015  &  & 0.198 & 0.411 & 0.390 && 6.0 & \it{6.16} \\
2000-2016 & & 0.198 & 0.401 & 0.401 && 6.79 & \it{8.72} \\
2000-2017 & & 0.187 & 0.443 & 0.370 && 12.02 & 8.43\\
\bottomrule
\end{tabular}
\label{tab1}
\end{table}

 
\begin{table}[t!]
\centering
\caption{\small \textbf{H-9.} Multi-step transition matrix probabilities for different training sets and MAPE for validation until 2019 of the Markov chain benchmark and Lotka-Volterra (LV) modeling. Italic numbers are used for LV whenever it yields a larger MAPE than the Markovian benchmark.}\vspace{0.5cm}
\begin{tabular}{cccccccc}
\toprule
 Years    & LULC & \multicolumn{3}{c}{Transition probabilities $P_{ij}$ \eqref{markov4}} & &\multicolumn{2}{c}{MAPE} \\
\cline{3-5} \cline{7-8}
 t$_i$-T$_T$   &  & Forest Plantations & Wetlands & Grasslands & & Markov & LV  \\ 
\hline
2000-2013 & Forest Plantations & 0.86 & 0.02 & 0.12 & & 11.63 & \it{14.25}
\\
 2000-2014 &  & 0.87 & 0.02 & 0.11 && 13.76 & 6.2 \\
2000-2015 &  & 0.88 & 0.02 & 0.10 && 15.97 & 1.46\\
 2000-2016 &  & 0.88 & 0.03 & 0.09 && 19.19 & 0.78 \\
2000-2017 & & 0.87 & 0.04 & 0.10 && 16.83 & 0.49\\
2000-2013 & Wetlands & 0.74 & 0.02 & 0.23 && 79.67 & 61.06 \\ 
2000-2014 &  & 0.77 & 0.02 & 0.21 && 80.42 & 15.28 \\
2000-2015 &  & 0.79 & 0.03 & 0.18 && 76.50 & 6.96  \\
2000-2016 & & 0.81& 0.04 & 0.16 && 73.36 & 13.10\\
2000-2017& & 0.79 & 0.05 & 0.16 && 61.12 & 9.11 \\
2000-2013 & Grasslands & 0.63 & 0.03 & 0.34 && 6.08 & \it{15.8} \\
2000-2014 &  &  0.66 & 0.03 & 0.31 && 12.12 & 9.46\\
2000-2015 &  & 0.70 & 0.04 & 0.27 && 21.42 & 7.47 \\
2000-2016 & & 0.73 & 0.04 & 0.23 && 34.00 & 9.71 \\
2000-2017 & & 0.72 & 0.06 & 0.22 && 31.09 & 13.69\\
\bottomrule
\end{tabular}
\label{tab2}
\end{table}
\begin{table}[b!]
\centering
\caption{\small $\overline{\mbox{MAPE}}$ for diverse training years in C-24 and H-9 regions.}\vspace{0.3cm}
\begin{tabular}{cccccc}
\toprule
 Years  &\multicolumn{2}{c}{C-24} & & \multicolumn{2}{c}{H-9} \\
\cline{2-3} \cline{5-6} 
  t$_i$-T$_T$  & Markov & LV & & Markov & LV\\
\hline
2000-2013 & 11.73 & 10.65 & &32.46 & 30.37 \\
2000-2014 &    8.17 & 7.37&& 35.43 & 10.31 \\
2000-2015 &   7.23 & 6.82 && 37.96 & 5.30 \\
2000-2016 &  10.46 & 10.17 & &42.18 & 7.86 \\
2000-2017 &  11.64 & 10.62 && 36.35 & 7.76 \\
\bottomrule
\end{tabular}
\label{tab3}
\end{table}

 \begin{table}[h!]
\centering
\caption{\small $\epsilon(t)$, the percentage area which is wrongly predicted for each year $t$ of the validation set  in C-24 and H-9 regions.}\vspace{0.5cm}
\begin{tabular}{ccccccc}
\toprule
 Years & $t$ &\multicolumn{2}{c}{C-24} & & \multicolumn{2}{c}{H-9} \\
\cline{3-4} \cline{6-7} 
t$_i$-T$_T$ && Markov & LV & & Markov & LV\\
\hline
2000-2013 &2014 & 8.09 & 5.33 & & 14.95 & 5.73 \\
 &2015 &    9.58 & 7.82 & & 13.68 & 13.23 \\
 &2016 &   11.78 & 12.48 & & 13.98 & \it{19.42} \\
 &2017 &  8.26 & 8.48 & & 16.61 & \it{19.61} \\
 &2018 &  5.87 & 8.87 & & 19.18 & \it{23.21} \\
 &2019 &  21.85 & 24.29 & & 19.90 & \it{28.62} \\
 \hline
2000-2014 &2015 & 4.35 & 2.50 & & 15.67 & 6.97 \\
 &2016 &    6.68 & 6.96 & & 16.93 & 9.76 \\
 &2017 &   3.94 & 3.05 & & 21.18 & 6.97 \\
 &2018 &  2.57 & 3.26 & & 23.89 & 7.02 \\
 &2019 &  16.74 & 19.04 & & 24.62 & 6.91 \\
 \hline
2000-2015 &2016 & 6.55 & 3.57 & & 16.64 & 3.71 \\
 &2017 &   3.49 & 3.87 & & 22.66 & 3.81 \\
 &2018 &  2.73 & 3.46 & & 25.67 & 3.01 \\
 &2019 &  17.00 & 12.70 & & 26.44 & 1.61 \\
 \hline
2000-2016 &2017 &  6.79 & 7.23 & & 22.88 & 2.28 \\
 &2018 &  5.03 & 4.85 & & 27.55 & 5.15 \\
 &2019 &  13.20 & 12.49 & & 28.56 & 3.53 \\
 \hline
2000-2017& 2018 &  4.48 & 4.25 & & 23.62 & 4.61 \\
 &2019 &  18.01 & 14.13 & & 26.41 & 2.69 \\
\bottomrule
\end{tabular}
\label{tab4}
\end{table}

\section*{Acknowledgements}\label{Ack}

\noindent Work partially supported by ANII through its SNI system and by  PEDECIBA-Uruguay.  J.A and J.A thanks A. Nahuel Lamas from the Universidad Tecnológica del Uruguay for its valuable support in the maps design. To the CYTED program under the grant number 520RT0010 Red GeoLIBERO. 
\newpage
\setcounter{figure}{0}
\renewcommand{\figurename}{Fig.}
\renewcommand{\thefigure}{S\arabic{figure}}

\begin{figure}[t!]
    \centering
    \includegraphics[scale=1.22]{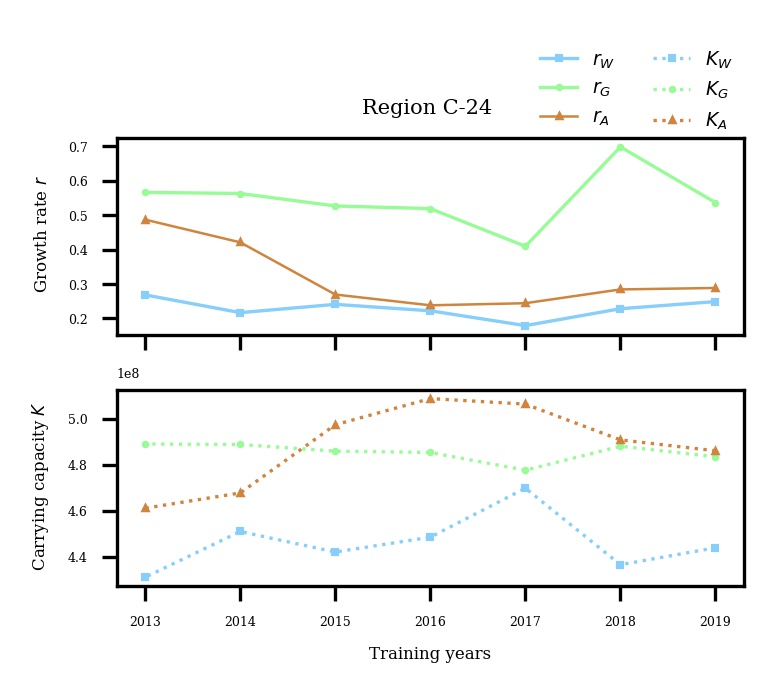} \hspace{-0.5cm}
    \includegraphics[scale=1.22]{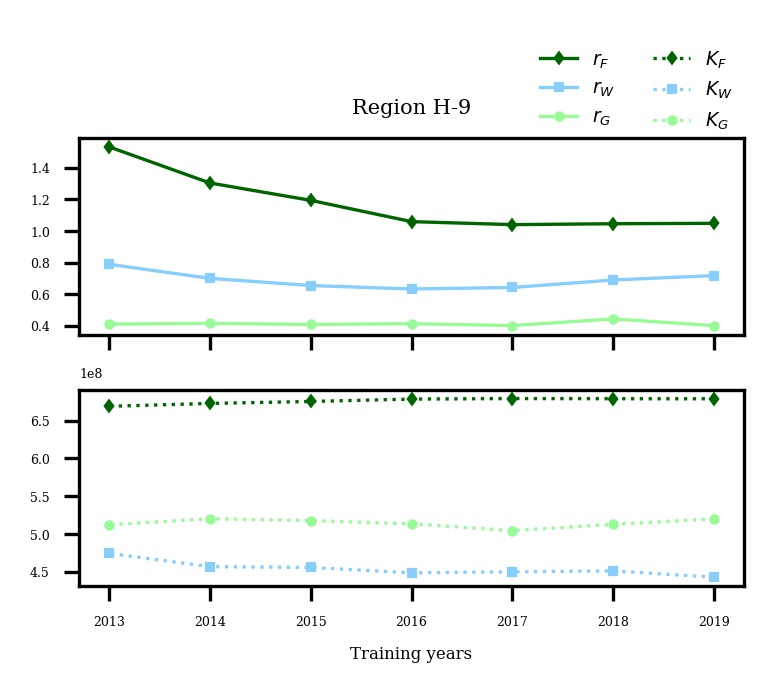}
    \caption{ Growth rate $r$ (solid lines) and carrying capacity $K$  (dashed lines)  for different training years (14-20 years since year 2000) obtained by linear regression. The sub-index $W,G,F,A$ corresponds to Wetlands, Grasslands, Forest Plantations and Agriculture, respectively.}
    \label{rk_c24}
\end{figure}


\bibliographystyle{unsrt}
\clearpage
\bibliography{main}

\begin{thebibliography}{10}

\bibitem{foley2005global}
Jonathan~A Foley, Ruth DeFries, Gregory~P Asner, Carol Barford, Gordon Bonan,
  Stephen~R Carpenter, F~Stuart Chapin, Michael~T Coe, Gretchen~C Daily,
  Holly~K Gibbs, et~al.
\newblock Global consequences of land use.
\newblock {\em science}, 309(5734):570--574, 2005.

\bibitem{diaz2019pervasive}
Sandra D{\'\i}az, Josef Settele, Eduardo~S Brond{\'\i}zio, Hien~T Ngo, John
  Agard, Almut Arneth, Patricia Balvanera, Kate~A Brauman, Stuart~HM Butchart,
  Kai~MA Chan, et~al.
\newblock Pervasive human-driven decline of life on earth points to the need
  for transformative change.
\newblock {\em Science}, 366(6471):eaax3100, 2019.

\bibitem{sala2000global}
Osvaldo~E Sala, FIII Stuart~Chapin, Juan~J Armesto, Eric Berlow, Janine
  Bloomfield, Rodolfo Dirzo, Elisabeth Huber-Sanwald, Laura~F Huenneke,
  Robert~B Jackson, Ann Kinzig, et~al.
\newblock Global biodiversity scenarios for the year 2100.
\newblock {\em science}, 287(5459):1770--1774, 2000.

\bibitem{pimm2000extinction}
Stuart~L Pimm and Peter Raven.
\newblock Extinction by numbers.
\newblock {\em Nature}, 403(6772):843--845, 2000.

\bibitem{gibbard2005climate}
Seran Gibbard, Ken Caldeira, Govindasamy Bala, Thomas~J Phillips, and Michael
  Wickett.
\newblock Climate effects of global land cover change.
\newblock {\em Geophysical Research Letters}, 32(23), 2005.

\bibitem{ROGAN2004301}
John Rogan and DongMei Chen.
\newblock Remote sensing technology for mapping and monitoring land-cover and
  land-use change.
\newblock {\em Progress in Planning}, 61(4):301--325, 2004.

\bibitem{graesser2015cropland}
Jordan Graesser, T~Mitchell Aide, H~Ricardo Grau, and Navin Ramankutty.
\newblock Cropland/pastureland dynamics and the slowdown of deforestation in
  latin america.
\newblock {\em Environmental Research Letters}, 10(3):034017, 2015.

\bibitem{jobbagy2006forestacion}
Esteban~G Jobb{\'a}gy, Mercedes Vasallo, Kathleen~A Farley, Gervasio
  Pi{\~n}eiro, Mart{\'\i}n~F Garbulsky, Marcelo~D Nosetto, Robert~B Jackson,
  and Jos{\'e}~M Paruelo.
\newblock Forestaci{\'o}n en pastizales.
\newblock {\em Agrociencia Uruguay}, 10(2):109--124, 2006.

\bibitem{paruelo2006cambios}
Jos{\'e}~M Paruelo, Juan~P Guerschman, Gervasio Pi{\~n}eiro, Esteban~G Jobbagy,
  Santiago~R Ver{\'o}n, Germ{\'a}n Baldi, and Santiago Baeza.
\newblock Cambios en el uso de la tierra en argentina y uruguay.
\newblock {\em Agrociencia Uruguay}, 10(2):47--61, 2006.

\bibitem{baldi2008land}
Germ{\'a}n Baldi and Jos{\'e}~M Paruelo.
\newblock Land-use and land cover dynamics in south american temperate
  grasslands.
\newblock {\em Ecology and Society}, 13(2), 2008.

\bibitem{modernel2016land}
Pablo Modernel, Walter~AH Rossing, Marc Corbeels, Santiago Dogliotti, Valentin
  Picasso, and Pablo Tittonell.
\newblock Land use change and ecosystem service provision in pampas and campos
  grasslands of southern south america.
\newblock {\em Environmental Research Letters}, 11(11):113002, 2016.

\bibitem{overbeck2007brazil}
Gerhard~E Overbeck, Sandra~C M{\"u}ller, Alessandra Fidelis, J{\"o}rg
  Pfadenhauer, Val{\'e}rio~D Pillar, Carolina~C Blanco, Ilsi~I Boldrini,
  Rogerio Both, and Eduardo~D Forneck.
\newblock Brazil's neglected biome: the south brazilian campos.
\newblock {\em Perspectives in Plant Ecology, Evolution and Systematics},
  9(2):101--116, 2007.

\bibitem{henwood2010toward}
William~D Henwood.
\newblock Toward a strategy for the conservation and protection of the world's
  temperate grasslands.
\newblock {\em Great Plains Research}, pages 121--134, 2010.

\bibitem{hoekstra2005confronting}
Jonathan~M Hoekstra, Timothy~M Boucher, Taylor~H Ricketts, and Carter Roberts.
\newblock Confronting a biome crisis: global disparities of habitat loss and
  protection.
\newblock {\em Ecology letters}, 8(1):23--29, 2005.

\bibitem{brazeiro2020agricultural}
Alejandro Brazeiro, Marcel Achkar, Carolina Toranza, and Luc{\'\i}a Bartesaghi.
\newblock Agricultural expansion in uruguayan grasslands and priority areas for
  vertebrate and woody plant conservation.
\newblock {\em Ecology and Society}, 25(1), 2020.

\bibitem{Lantman2011}
Jonas van Schrojenstein~Lantman, Peter~H. Verburg, Arnold Bregt, and Stan
  Geertman.
\newblock {\em Core Principles and Concepts in Land-Use Modelling: A Literature
  Review}, pages 35--57.
\newblock Springer Netherlands, Dordrecht, 2011.

\bibitem{Baker89}
William~L. Baker.
\newblock A review of models of landscape change.
\newblock {\em Landscape Ecology}, 2:111–133, 1989.

\bibitem{lambin1997}
Eric~F. Lambin.
\newblock Modelling and monitoring land-cover change processes in tropical
  regions.
\newblock {\em Progress in Physical Geography: Earth and Environment},
  21(3):375--393, 1997.

\bibitem{burnham_1973}
Bruce~O. Burnham.
\newblock Markov intertemporal land use simulation model.
\newblock {\em Journal of Agricultural and Applied Economics}, 5(1):253–258,
  1973.

\bibitem{kumar2013}
Sathees Kumar, Nisha Radhakrishnan, and Samson Mathew.
\newblock Land use change modelling using a markov model and remote sensing.
\newblock {\em Geomatics, Natural Hazards and Risk}, 5(2):145--156, 2014.

\bibitem{apellaniz2021temperate}
Melisa Apellaniz, Niall~G Burnside, and Matthew Brolly.
\newblock Temperate grassland afforestation dynamics in the aguapey valuable
  grassland area between 1999 and 2020: Identifying the need for protection.
\newblock {\em Remote Sensing}, 14(1):74, 2021.

\bibitem{ozturk2015urban}
Derya Ozturk.
\newblock Urban growth simulation of atakum (samsun, turkey) using cellular
  automata-markov chain and multi-layer perceptron-markov chain models.
\newblock {\em Remote Sensing}, 7(5):5918--5950, 2015.

\bibitem{yang2012spatiotemporal}
Xin Yang, Xin-Qi Zheng, and Li-Na Lv.
\newblock A spatiotemporal model of land use change based on ant colony
  optimization, markov chain and cellular automata.
\newblock {\em Ecological Modelling}, 233:11--19, 2012.

\bibitem{girma2022land}
Rediet Girma, Christine F{\"u}rst, and Awdenegest Moges.
\newblock Land use land cover change modeling by integrating artificial neural
  network with cellular automata-markov chain model in gidabo river basin, main
  ethiopian rift.
\newblock {\em Environmental Challenges}, 6:100419, 2022.

\bibitem{pastor2008mathematical}
John Pastor.
\newblock {\em Mathematical ecology of populations and ecosystems}.
\newblock John Wiley \& Sons, 2008.

\bibitem{fort2020ecological}
Hugo Fort.
\newblock {\em Ecological Modelling and Ecophysics}.
\newblock IOP Publishing, 2020.

\bibitem{10.1371/journal.pcbi.1004182}
Richard~R. Stein, Debora~S. Marks, and Chris Sander.
\newblock Inferring pairwise interactions from biological data using
  maximum-entropy probability models.
\newblock {\em PLOS Computational Biology}, 11(7):1--22, 07 2015.

\bibitem{IGM}
Instituto geográfico militar - uruguay.
\newblock \url{https://igm.gub.uy/}.

\bibitem{mapbiomas}
Mapbiomas-pampa project.
\newblock \url{http://pampa.mapbiomas.org}.

\bibitem{jaynes1957information}
Edwin~T Jaynes.
\newblock Information theory and statistical mechanics.
\newblock {\em Physical review}, 106(4):620, 1957.

\bibitem{jaynes1957information2}
Edwin~T Jaynes.
\newblock Information theory and statistical mechanics. ii.
\newblock {\em Physical review}, 108(2):171, 1957.

\bibitem{shannon1948mathematical}
Claude~Elwood Shannon.
\newblock A mathematical theory of communication.
\newblock {\em The Bell system technical journal}, 27(3):379--423, 1948.

\bibitem{stein2015inferring}
Richard~R Stein, Debora~S Marks, and Chris Sander.
\newblock Inferring pairwise interactions from biological data using
  maximum-entropy probability models.
\newblock {\em PLoS computational biology}, 11(7):e1004182, 2015.

\bibitem{HFort}
Hugo Fort.
\newblock {\em Ecological Modelling and Ecophysics}.
\newblock 2053-2563. IOP Publishing, 2020.

\bibitem{fort2021method}
Hugo Fort and Tom{\'a}s~S Grigera.
\newblock A method for predicting species trajectories tested with trees in
  barro colorado tropical forest.
\newblock {\em Ecological Modelling}, 446:109504, 2021.

\bibitem{fort2021new}
Hugo Fort and Tom{\'a}s~S Grigera.
\newblock A new early warning indicator of tree species crashes from effective
  intraspecific interactions in tropical forests.
\newblock {\em Ecological Indicators}, 125:107506, 2021.

\bibitem{emary2021markets}
Clive Emary and Hugo Fort.
\newblock Markets as ecological networks: inferring interactions and
  identifying communities.
\newblock {\em Journal of Complex Networks}, 9(2):cnab022, 2021.

\bibitem{wang2021land}
Sonam~Wangyel Wang, Lamchin Munkhnasan, and Woo-Kyun Lee.
\newblock Land use and land cover change detection and prediction in bhutan's
  high altitude city of thimphu, using cellular automata and markov chain.
\newblock {\em Environmental Challenges}, 2:100017, 2021.

\bibitem{WENG2002273}
Qihao Weng.
\newblock Land use change analysis in the zhujiang delta of china using
  satellite remote sensing, gis and stochastic modelling.
\newblock {\em Journal of Environmental Management}, 64(3):273--284, 2002.

\bibitem{wetlands}
Liwei Xing, Huabin Wang, Wenfeng Fan, Chen Chen, Tao Li, Guanghui Wang, and
  Haoran Zhai.
\newblock Optimal features selection for wetlands classification using landsat
  time series.
\newblock In {\em IGARSS 2018 - 2018 IEEE International Geoscience and Remote
  Sensing Symposium}, pages 8385--8388, 2018.

\bibitem{wet2}
Timothy~G. Whiteside and Renée~E. Bartolo.
\newblock Mapping aquatic vegetation in a tropical wetland using high spatial
  resolution multispectral satellite imagery.
\newblock {\em Remote Sensing}, 7(9):11664--11694, 2015.

\bibitem{prices}
Pulp, paper, and allied products prices.
\newblock \url{https://fred.stlouisfed.org/series/WPU0911}.

\bibitem{rs}
PETER~H. VERBURG, KATHLEEN NEUMANN, and LINDA NOL.
\newblock Challenges in using land use and land cover data for global change
  studies.
\newblock {\em Global Change Biology}, 17(2):974--989, 2011.

\end{thebibliography}
\end{document}